\documentclass{article}
\usepackage[dvips]{graphics}
\usepackage{latexsym,epsfig, amssymb}
\author{M.~Meier-Schellersheim, G.~Mack\\
  II.Institut f\"ur Theoretische Physik\\
  Universit\"at Hamburg\footnote{email
    martin.meier-schellersheim@desy.de, mack@lienhard.desy.de}\\
}

\begin{document}

\title{SIMMUNE,\\ a tool for simulating and analyzing\\
  Immune System behavior 
\footnote{Work supported by Deutsche Forschungsgemeinschaft and by the German
  Israeli Foundation for Scientific Research and Development (GIF)}}

\maketitle

\centerline{\bf Abstract}

We present a new approach to the simulation and analysis of immune system (IS)
 behavior. 
The simulations that can be performed with our software package called 
SIMMUNE are based on immunological data that describe the behavior of 
IS agents (cells, molecules) and the IS's challengers (bacteria, viruses) 
 on a microscopical (i.e. agent-agent interaction) scale by defining cellular 
stimulus response mechanisms.
All processes within the simulated IS are based on these mechanisms.
Since the behavior of the agents in SIMMUNE can be very flexibly
configured, its application is not limited to IS simulations.\\ 
We outline the principles of SIMMUNE's multiscale 
analysis of emergent structure within the simulated IS that allow the 
identification of immunological contexts using minimal {\it a 
priori} assumptions about the higher-level organization of the IS.\\
{\bf Keywords}: immune system simulation, locally interacting agents, 
multiscale analysis.\\

\section{Introduction}

For quite a long time immunological research limited itself mainly to the 
investigation of molecular details of cellular mechanisms within the immune
system (IS). 
The structure of immune responses was believed to be rather simple: 
Upon infection of the organism IS agents had to 'recognize' the foreign
material that caused the infection, the \emph{pathogen}, with highly
specific receptors and then remove everything within the organism
that would bind to these receptors. To make this principle work, the IS
only had to ensure that it did not produce receptors complementary to
components of host organism.\\
From this point of view, developing a vaccine against some kind of disease 
thus simply meant searching for the right (harmless) fragment of the 
pathogen that could be presented to the IS as an \emph{antigen} (a substance 
that provokes an immune response). Having recognized this fragment once,
the IS would memorize this knowledge through maintaining a large enough
number of the right receptors and then upon contact with the real
 pathogen roll out its arsenal of defense quickly enough to avoid a
 spreading of the pathogen within the organism.\\
In many cases vaccine development is more complicated.
The IS normally looks for more than just one signal (the antigen) before
it produces a full response. Given that the IS can hardly rule out
every receptor that fits to material of the own organism without crippling
its own functionality because of the degeneracy of receptors
\footnote{IS receptors usually fit to more than just one single molecular
structure. The neccessity to be able to provide receptors to virtually
any (foreign) molecular structure the IS might get confronted to, prevents
the IS from destroying every receptor that fits to the material of its
host organism.}   
it is understandable that it should ask for more information.\\
(For a mathematical discussion of the problem of immune receptor 
specificity see for example \cite{PERELSON}.)\\ 
Providing the right \emph{adjuvans}, the biochemical context within
which the antigen is seen by the IS, is important. Adjuvants often contain
fragments of classical pathogens, i.e. pathogens the organism and its
ancestors have been accustomed to for a long time. Being confronted to
 these pathogen fragments the IS more readily switches into 'defense mode'.\\
Unfortunately, within certain contexts the IS readily accepts the
 organism's own material as antigenic. Autoimmune diseases like
 diabetes are examples where the IS, once it finds itself in a certain
 context, attacks its own host.\\

\noindent Cohen \cite{IRUN1} formulated the \emph{cognitive paradigm} 
postulating that cognitive abilities - enabling the IS to select its response
according to the context of the presentation of antigen - were an
indispensable ingredient of IS behavior.\\
Grossman \cite{GROSS} explained why we should investigate the IS's
context recognition and proposed some models for appropriate
cell behavior.\\
Segel and Bar-Or \cite{SEGEL1} investigated the question how the IS's 
cells might be able to optimize their contribution to immune responses 
with the help of feedback mechanisms and compared the IS to other 
systems of decentralized organizing agents. Segel \cite{SEGEL2} also presented
the idea of a \emph{diffuse informational network} of cytokines encoding the 
molecular context the IS finds itself in. Atlan and Cohen \cite{IRUN3} pointed out 
that the IS must achieve the ability to extract -- in a cognitive process --
\emph{meaning} from the wealth of \emph{information} its cells gather via 
their receptors.\\  
Often, the term 'recognition' in an immunological sense refers to the ability
of the IS to provide receptors that can bind to the surface of 
foreign material with high enough affinity to direct an effective
immune response against this material. Recognition of this kind 
comes down to the question whether molecular shapes are complementary
while the ability to act context dependent requires the IS's agents
to mutually coordinate the processing of the molecular signals
which they receive from their milieu.\\
How could the IS's information processing work to achieve context
recognition? What is the nature of the context, that make the IS respond in
a protective (or harmful) manner? How is the IS provided
with the specific context that would activate an immune response to 
some threatening disease?\\
To be able to start answering these questions, we need to investigate
how the agents of the IS exchange information, how they influence
each other's states, how their reactions to combined signals differ
from their reactions to these signals when occurring at different
times. Further, what is the spatial scope of specific signals, {\it i.e.}, do they
influence only their direct neighborhood or do they spread over larger
areas of the organism?\\

\noindent The software package SIMMUNE which we want to introduce here
was designed to facilitate simulations dealing with these questions.\\ 
SIMMUNE simulates the IS on the agent level, {\it i.e.} on the level of
interactions between cells and molecules. The analysis that is performed on 
the simulations, however, operates on multiple scales; this is described in 
section 3.4.
Application of the approach of identifying immunological
contexts by multiscale analysis to more elaborate simulations will be described
in a forthcoming publication \cite{MMS1}.\\
 
\noindent In section 2 we will briefly describe some of the classical methods of 
IS modelling to be able to point out the differences and similarities 
between them and the approach presented here.
In section 3 our software package SIMMUNE will be introduced. We will
give an overview over its structure and present some simple
examples of SIMMUNE applications, as well as some of the methods to
analyze the simulations. Section 3 concludes with some remarks
on the limitations of the current version of SIMMUNE.\\
In section 3.3.2 we will describe some immunological mechanisms as far as
they are implemented in the example simulations. 
Descriptions of immunological mechanisms refer to the way they are implemented in 
the simulations which we present here. In those cases where they differ
fundamentally from their real biological counterparts we will mention this.

\section{Modelling IS Behavior}

The possibilities of tracing directly the complex sequences of interactions
in real ISs of living organisms are very limited. However, Jenkins \cite{JENK} 
has presented a method of monitoring the activities of selected cell clones
\emph{in situ}, i.e. in the living organism. (A definition 
of 'cell clones' as they are implemented in our simulations will be given 
in chapter 3.3 .)\\ 
Various techniques of IS modelling have been developed that allow to 
investigate theoretically different aspects of immunology.
Perelson and Weisbuch \cite{PERELSON} have provided a comprehensive survey of 
this area.

\subsection{Reaction-Kinetics Models}
The great complexity of IS behavior is due to the large number of 
different types of IS agents (cells, molecules) that can interact with
each other in various ways. Deriving the system's behavior from the
interactions between its many constituents is one of the goals of 
theoretical immunology.\\
Abstracting from the specific details of interaction between the IS's 
agents one can formulate systems of coupled differential equations 
describing how the time development of an agent's concentration 
in the modelled organism depends on the concentrations of other types 
of agents. We call these models \emph{reaction-kinetics} models 
as they bear resemblence to models of reaction kinetics in 
chemistry. A very simple example for such a system is the following.

\begin{eqnarray*}
&\frac{dI}{dt}& = p_{infect} I C - p_{kill}IK - d_{I}I\\
&\frac{dK}{dt}& = p_{resp}IK - d_{K}K\\
&\frac{dC}{dt}& = s - p_{infect}IC - d_{C}C
\end{eqnarray*}

\begin{figure}[h]
\begin{center}
\includegraphics[width=6cm,angle=270,clip=]{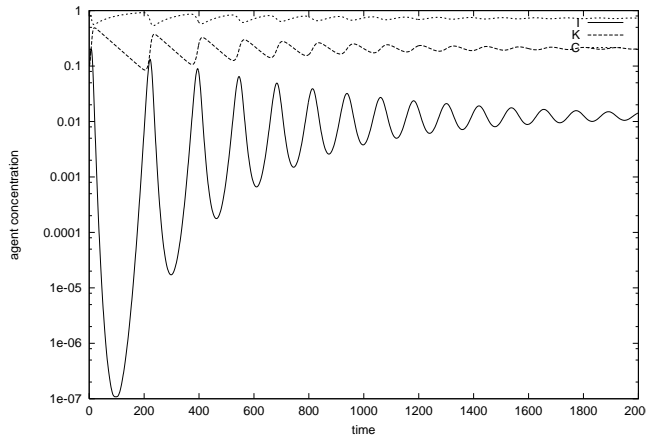}
\end{center}
\caption{The model IS settles down into 'cronic infection' after oscillations
with very low concentrations of {\it I}.}
\label{osclog}
\end{figure}

Using as a shortcut notation \emph{I}, \emph{K} and \emph{C} for agent 
names as well as for
their concentrations, the equations above describe a situation that may be 
interpreted as follows: Infectious agents
of type \emph{I} transform agents of type \emph{C} into new agents of
type \emph{I} upon contact with a rate $p_{infect}$.
Agents of type \emph{I} get removed (killed) upon contact with agents of type
\emph{K} with a rate $p_{kill}$. Agents \emph{K} proliferate upon contact 
with \emph{I} with a rate $p_{resp}$.
Agents of all three types die naturally at their specific rates $d_{I}$,  
$d_{K}$, $d_{C}$. \emph{C} type agents are produced at a constant 
rate \emph{s}.
\begin{figure}[h]
\begin{center}
\includegraphics[width=6cm,angle=270,clip=]{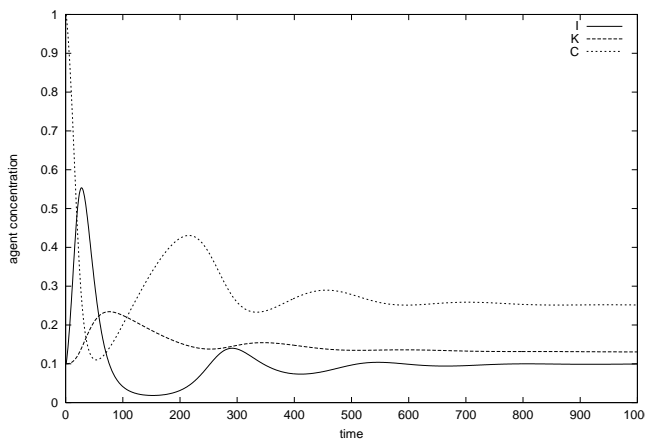}
\end{center}
\caption{time development of agent concentration in the simple IS model}
\label{cro}
\end{figure}
\emph{K} could be considered to be an immune system cell type being produced 
as a response to the appearance of the infectious \emph{I}. \emph{C} may
be presenting any possible type of target cell for the pathogen \emph{I}.\\
Integrating this system of equations yields different kinds of time development
for \emph{I}, \emph{K} and \emph{C} depending on the parameters $p_{infect}$, 
$p_{kill}$, $p_{resp}$, \emph{s}, the death rates $d_{x}$ and 
of course the initial values of \emph{I}, \emph{K} and \emph{C}.\\
Fig. \ref{cro} shows the time development of the agents for one set of
parameters (see appendix). We see how at first the number of infected agents 
\emph{I} grows while the number of
'healthy' agents \emph{C} decreases. Then, as the response from the IS 
agents \emph{K} grows, the number of infected agents declines while \emph{C}
recovers. Finally the system ends up in a steady state that may be interpreted
as a chronic infection: Infectious agents persist even though the IS constantly 
prepares agents of type \emph{K} to suppress the infection.\\
Other sets of parameters may lead to stronger oszillations of the agents 
concentrations.
Fig. \ref{osclog} shows in a half-logarithmic plot a system that settles down into a
steady 'chronic infection' state after having gone through states where the 
concentration of \emph{I} is very low. 

Fig. \ref{osclog} illustrates one of the limits of applicability of simulation 
results of the \emph{reaction-kinetics} approach: In a real system of 
interacting entities at a given time all agents of a certain type may have 
dissappeared due to destructive interaction with other agents (in our 
example, such an interaction is for example the suppression of \emph{I} 
by \emph{K}).
In the \emph{reaction-kinetics} approach there are no individual agents.
Even agent types with unrealistically low concentrations may experience a 
'comeback'.
Of course, this problem can be avoided by appropriate setup of the model,
but it points out one of the weaknesses of the approach: \emph{Information}
is \emph{global} in \emph{reaction-kinetics} models. In contrast, information in
nature is something \emph{local}.\\
In physics, this principle of \emph{locality}
is one of the major foundations of modern theories. It lies at the heart of
field theory and can be used as a starting point for entering Einstein's
General Theory of Relativity.\\
In immunology, the locality of information processing by the IS's agents
may prove to play an important role too.
   
\subsection{Automata}
Automata models of the IS neglect the microscopic details of IS behavior. 
They identify a set of characteristic states of the IS (like 'in rest' or
'with infection') together with transition
rules that define how the automaton may switch from one state to another.\\
At discrete timesteps the state of the automaton is evaluated and the rules
are applied to define the automaton's state for the next timestep.\\
Atlan and Cohen \cite{ATLAN} investigated the effects of suppressor T cells
using a neural network automaton.

\subsection{Cellular Automata}
Cellular automata (CA) were invented to investigate how simple building blocks
could locally cooperate to produce aggregates with interesting behavior. 
The building blocks are automata living on a grid. Their rules define how the 
change in state of a single automaton at the next time step depends on his own 
current state and the states of his direct neighbor automata.\\
Clearly, the most fascinating aspect of CA modelling is the fact that even 
rather simple transition rules together with strictly \emph{local} interactions 
can lead to very complex behavior of automata aggregates. An overview and 
classification of different types of CAs can be found in \cite{WOLF}.\\ 
In immunology we encounter a similar situation -- all IS activities are
based on the actions of cells reacting to their direct neighbor cells and 
molecules. There is no central supervision of immune responses.
Nevertheless the IS manages to coordinate the actions of its constituents
over larger spans of space and time.\\

\noindent Celada and Seiden \cite{SEID1} developed a CA model of the IS where the 
cells are simulated by automata that may (as a modification of the usual CA
concept of static correspondence between information and position) carry their
state information with them while they are moving on a 2-dimensional grid. 
Depending on the agents (cells, molecules) they encounter they may change 
their state \emph{e.g.} from \emph{naive} to \emph{activated}.\\
With their simulation program called IMMSIM Celada and Seiden were 
able to investigate a number of IS phenomena, for example \emph{Affinity 
maturation and hypermutation [\ldots] of the humoral immune response} 
\cite{SEID2}.

\section{SIMMUNE}
\begin{figure}[h]
\begin{center}
\includegraphics[bbllx=260,bblly=410,bburx=540,bbury=785,width=6cm,angle=90,clip=]{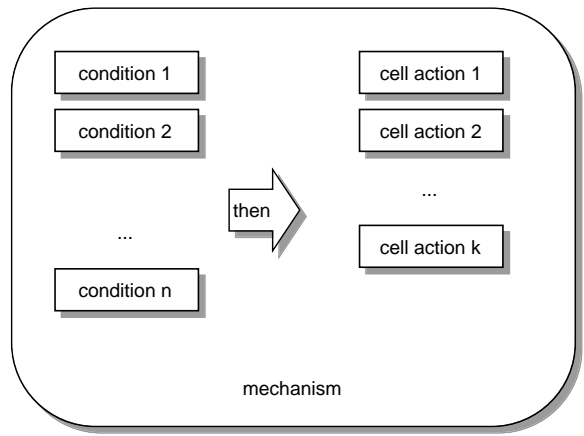}
\end{center}
\caption{cellular mechanism}
\label{mechanism}
\end{figure}

SIMMUNE is an attempt to derive IS behavior from immunological data that
describe the behavior of the cells of the IS on a microscopical level
by defining cellular stimulus response mechanisms.\\
A cellular stimulus response mechanism (\emph{cellular mechanism} for short)
consists of a description of a set of stimuli a cell needs to experience
before it performs certain actions, and a description of those actions.\\
Metaphorically, a mechanism thus may be considered to be a set of conditional
actions: The cell checks whether certain conditions are fulfilled and if
they are, it performs certain actions.\\
In SIMMUNE, the condition part of a mechanism can consist of an arbitrary 
number of conditions that can be combined through logical \emph{AND} or 
\emph{AND NOT}. The action part may consist of one or more actions.
Fig. \ref{mechanism} illustrates the concept of a cellular mechanism.
\begin{figure}[h]
\begin{center}
\includegraphics[bbllx=189,bblly=125,bburx=578,bbury=800,width=7cm,angle=90,clip=]{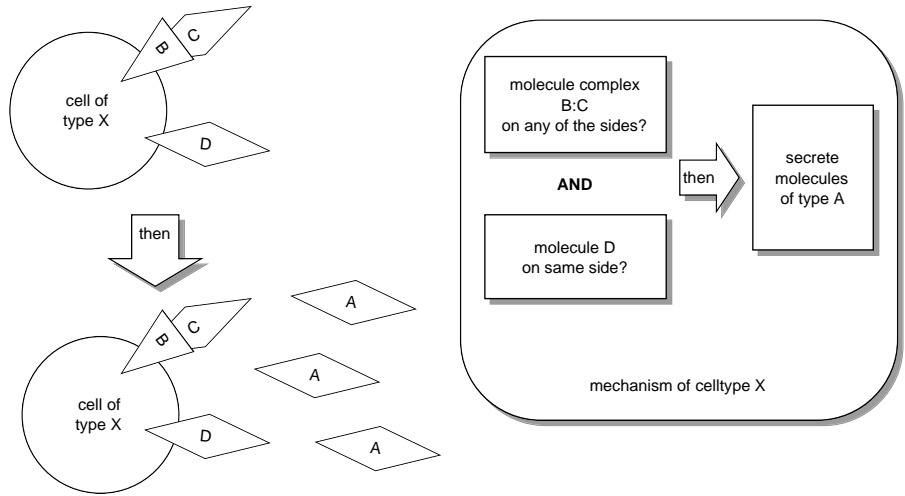}
\end{center}
\caption{an example of a cellular mechanism}
\label{examplemechanism}
\end{figure}

Fig. \ref{examplemechanism} presents an example of a cellular
mechanism. There, the stimuli consist of the \emph{B:C} complex and the 
\emph{D} molecule on the cell's surface.

\subsection{Components of SIMMUNE}
\subsubsection{Celltypes}
Like in cellular automata models, in SIMMUNE too cells are 
individual entities that interact with each other only locally.
The cells live on a 3-dimensional grid. Each cell in a SIMMUNE model 
belongs to a certain cell type that is defined by the set of 
mechanisms according to which cells of a this type act. SIMMUNE cell types
need not be equivalent to cell types as they are identified in experimental
immunology.\footnote{Interestingly enough the categorization of IS cells
into rigidly distinct cell types has often been a controversial subject in
immunology. Currently this is the case for T-helper1 vs. T-helper2 cells.}\\
It is important to notice that mechanisms in SIMMUNE do not define cell 
\emph{states} (that would be described by a fixed set of attributes, like
certain types of molecules on the cell's surface) but cell \emph{behavior},
while the usual approach in CA immune system 
models is to define a set of cell states and the rules how cells may 
switch from one state to another (cellular actions simply being the process
of changing the cell's state). \\
Depending on the stimuli they receive from their environment, cells in 
SIMMUNE may change their attributes in various ways according to the mechanisms 
of their type.
They may express receptor molecules on their surface, incorporate
material from the extracellular milieu, secrete certain messenger substances,
kill neighbor cells or move into a certain direction. They may also have
part of their mechanisms modified (for example as a consequence of viral
attack) without completely changing their type.\footnote{
Here, the analogy between sets of cellular mechanisms in SIMMUNE and the
  genetic encoding of cellular behavior in real biological systems is
obvious.}

\begin{figure}[h]
\begin{center}
\includegraphics[bbllx=110,bblly=455,bburx=570,bbury=815,width=9cm,angle=90,clip=]{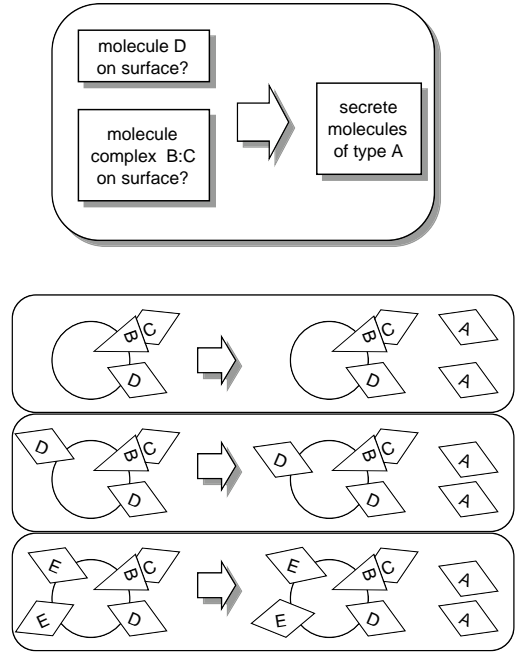}
\end{center}
\caption{different ways of satisfying common conditions}
\label{differentways}
\end{figure}

In SIMMUNE several conditions, each checking a simple attribute of the 
cell's state may be combined in the condition part of a mechanism.
A condition for a given cell action may be fulfilled in many different 
ways -- all of which may bear specific information for other mechanisms.\\
Fig. \ref{differentways} illustrates how the mechanism based approach of SIMMUNE is able to
provide more flexibility in describing cell behavior than IS models that 
completely define cell states and transitions between them.\\

\noindent Besides its mechanisms, cell type properties in SIMMUNE 
include typical (mean) lifetime and the size of the cells.\\

\subsubsection{Molecules}
\noindent In SIMMUNE cells are the basic units of signal- and information processing.
The signals they receive consist of molecules from the extracellular milieu 
that bind to the receptor molecules which the cells display on their surfaces.
(Molecules also perform signal- and information processing
simply by forming aggregates according to their mutual binding possibilities.
In real ISs certain kinds of molecular aggregates are responsible of attacking
cellular pathogens as a first line of defense before the IS has build up its 
complete arsenal of effector cells. 
The death of a cellular pathogen as a reaction to contact with
the above mentioned molecular aggregates would, however, be encoded as a
\emph{cellular} action in SIMMUNE.)
The significant properties of a molecule hence are described by the set of its
possibilities to bind to other molecules.\\ 
A part of a molecule that
is visible to the milieu and hence can be used to bind the molecule to
receptors is called an \emph{epitope}. Even though one usually only
refers to the binding sites of 'normal' molecules as epitopes, we treat 
receptor binding sites just as 'normal' molecule binding sites and call them 
epitopes too.\\
The binding possibilities of any molecule are thus defined by the binding
possibilities of its epitopes.\\  
Other molecule properties in SIMMUNE include the typical lifetime of the 
molecule before it desintegrates into its fragment molecules and the types 
of fragments that are produced upon desintegration of the molecule.

\subsubsection{Compartments}
The effects of many interactions in the IS are local by their nature. An
example is the cell-cell communication via contact receptors.
But moving molecules or cells may influence all those parts of the organism
which they can access.\\
To provide the right specific milieus for the different tasks it has
to fulfill, the IS uses different compartments.
For example,  as mentioned above, the
IS needs to avoid producing too many cells with receptors with high 
affinity for material of the own organism. It achieves this by establishing
a central 'school' compartment, the \emph{thymus}, where cells of a 
certain very important IS celltype, the \emph{T-cells}, are tested to 
be useful and to be not too self-reactive before they are allowed to
start their immune activities.\\
Another wellknown type of IS
compartment are the \emph{lymph nodes}. They are used to bring together
the different cell types of the IS for information exchange.\\
Every type of IS cell may be considered to see a different aspect of an antigen.
Hence, in order to exploit all the information available about the antigen, 
the IS needs to provide the lymph nodes as meeting points for the 
different cells carrying different pieces of information.
From a simplified and abstract point of view, compartments simply gather 
certain kinds of agents while excluding others.
A simulation with SIMMUNE may comprise different compartments.

\subsection{Operation of SIMMUNE}
Besides the properties of cell and molecule types, SIMMUNE lets its
user define the properties of the compartments within the simulated IS. 
Dimensions of the compartments,
diffusion rates of molecules and cells within the compartments as well as
initial concentrations
of the different types of agents can be given. Furthermore, the exchange
of agents between the different compartments can be regulated: which kinds
of agents are allowed to pass from one compartment to another and at which rate.\\
SIMMUNE offers a graphical user interface that can be used to watch agents'
concentrations and the spatial distribution of cells or manipulate the running
simulation by injecting new cells or molecules.

\subsection{Example Applications}
\subsubsection{Local vs. Global Interaction between Agents}
\begin{figure}[h]
\begin{center}
\includegraphics[bbllx=175,bblly=430,bburx=565,bbury=815,width=8cm,angle=90,clip=]{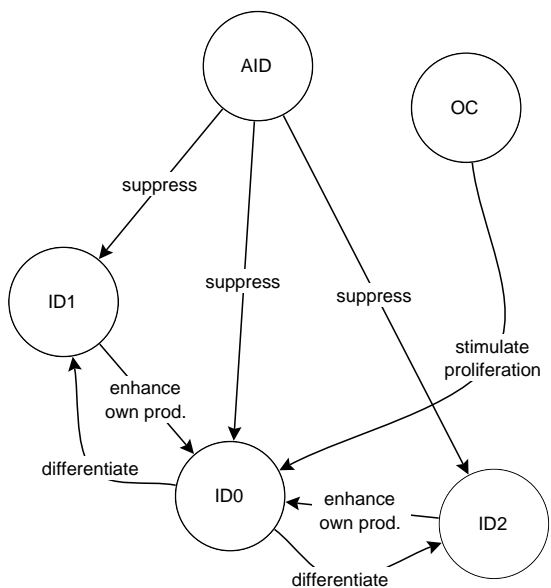}
\end{center}
\caption{Network of cell interactions in the 'Local vs. Global Interaction' simulation}
\label{feedback}
\end{figure}

This example is meant to demonstrate with a very simple model 
differences between a \emph{reaction-kinetics} simulation and a
simulation with locally interacting agents. Five types of cells interact 
as illustrated in Fig. \ref{feedback}.\\
\emph{ID0} cells proliferate upon contact with cells of type \emph{OC}. 
Upon contact with certain \emph{cytokines} (signal molecules) they transform 
into cells of type \emph{ID1} or \emph{ID2}, depending
on the kind of cytokine they register. \emph{ID1} and \emph{ID2} themselves produce
the cytokines ${C}_{1}$ and ${C}_{2}$ that make their precursor $ID0$ switch 
into their
own state respectively. \emph{AID} cells kill all \emph{ID} type cells upon contact.

\noindent Translation of this interaction network into a system of coupled 
differential
equations leads to a situation with an unstable equilibrium between the
two cell types \emph{ID1} and \emph{ID2}. Because of the positive feedback between
\emph{ID1/ID2} and \emph{ID0} small deviations from equal concentrations
of both types may push the system into nearly exclusive production of
one type.\\

\noindent Simulation of this system with locally interacting agents yields a different
behavior. The cytokines that are secreted by \emph{ID1} and \emph{ID2} do not instantly
spread all over the compartment. If the cytokines' lifetimes are
short and diffusion is not too strong, \emph{ID1} and \emph{ID2} act with their
cytokines only on \emph{ID0} cells that are located within a small neighborhood of 
themselves.\\ 
The parameters of the simulation are given in the appendix.\\
Fig. \ref{symslice} in the appendix shows a cut through the compartment after 
the simulation has reached a state of dynamic equilibrium.
Neither \emph{ID1} nor \emph{ID2} have achieved global dominance. Instead, the two
competitors cluster in areas with a diameter of typically 5-10 cells.
Another interesting effect is that the \emph{AID} cells do not appear
inside of these clusters. Although they perform free random moves (controlled
only by the availability of free space) they remain on the clusters'
surfaces. One is reminded of the situation of a water drop on a hot stove: 
while the cells on the surface of the cluster get killed, those cells inside
the cluster manage to survive.

\subsubsection{A simple Immune System}
Here we demonstrate a simple IS model simulation. The simulation
includes five types of cells, three of which may go through
differentiations during their lifetimes. Furthermore one type of 
virus is part of the model.\\
Most of the cellular mechanisms in this model are strong simplifications
of the processes in real ISs. The reason why we present this simulation
is that all of its behavior results from direct cell-cell respectively
cell-molecule interactions and that it suffices as an example simulation 
that allows to describe how we analyze the behavior of the simulations 
and to explain the notion of an IS \emph{context}. We will return to
this in section 3.4 .\\
 
\noindent The first cell type \emph{OC} is an organism cell not belonging to the IS.
Cells of this type divide at a certain (low) rate. This reproduction is
controlled by \emph{contact inhibition}: \emph{OC} cells have receptors on their
membranes that are able to bind to receptors of the same type on the 
membranes of neighbor cells. If such a cell finds too many of these receptor
complexes on its surface, it refrains from dividing.\\
\emph{OC} cells are the target of the virus \emph{V}. Upon contact with
\emph{V} the \emph{OC} type cells get infected and start producing new viruses that
are kept inside the cells. After some time however the cells burst and
release their virus content to the extracellular milieu. Even though
infected cells do not constitute a completely new cell type in immunological
terminology they will
be called \emph{IC} type cells here to facilitate the discussion of the model
and to identify them in the diagrams.\\
\emph{IC} cells besides producing new viruses present virus epitopes on 
\emph{MHC1} receptor molecules on their surface. They do not proliferate
and have a shorter lifespan than \emph{OC} cells.\\

\noindent T-Cells appear in three states of activity. As naive T-cells \emph{NT} and 
activated T-killer (\emph{TK}) or T-helper (\emph{TH}) cells. 
\emph{NT} cells possess two mechanisms. The first makes them 
express receptor molecules \emph{TCR} on their surfaces, the second mechanism
induces their transformation into \emph{TK} or \emph{TH} cells and proliferation 
(reproduction). Using a rather simplified T-cell activation scheme, we define
the stimulus for this transformation to be the presence of 
a \emph{TCR} being bound to a receptor molecule of type \emph{MHC1} that besides its 
own epitopes presents an additional antigen epitope.\\
A receptor molecule \emph{TCR} possesses a binding site that uses two epitopes.
Both of these are selected at random. This means their binding possibilities
(i.e. which epitopes they can bind to) are arbitrarily chosen. Each \emph{NT} cell
once in its lifetime selects its own (random-) epitopes. All \emph{TCR} molecules 
produced by this cell will possess these two epitopes. When the cell transforms
into a \emph{TK} or \emph{TH} cell it keeps this epitope choice; if such a cell
divides, both daughtercells also will use these two epitopes for their
receptors. Thus all \emph{TK} cells stemming from a common \emph{NT} cell will bind
with their receptors to the same \emph{MHC1}/antigen complexes. These cells are said
to constitute a \emph{clone}.\\
A \emph{TCR} to be able to bind to a \emph{MHC} must be able to bind to the 
\emph{MHC}'s
own epitopes as well as to the antigen epitope which the \emph{MHC} presents.\\
Naive T-cells (\emph{NT}) are stimulated to change their type and become
(activated) \emph{TK} or \emph{TH} cells upon registering a \emph{TCR:MHC} complex. 
\emph{TK} cells react to such complexes by killing the cell that presents 
the \emph{MHC} receptor. This is the way the IS tries to remove infected
cells (\emph{IC}) before they are able to release their virus load to the milieu.\\
The \emph{TCR}s of \emph{TH} cells bind not to \emph{MHC1} but to similar membrane molecules
called \emph{MHC2} that are used by B-cells (see below).
Similar to the contact inhibition of \emph{OC} cells, \emph{TK} and \emph{TH}
cells have a contact mediated mechanism that controlls their proliferation. 
\footnote{Knowledge about the mechanisms that are used by \emph{real} ISs to controll
  T-cell proliferation seems to be still rather limited.} They have receptors (called \emph{FAS}) and their counterparts
(\emph{FAS-ligand}) 
on their surface. An activated T-cell that finds a \emph{FAS:FAS-ligand} complex on its
surface commits suicide.\\

\noindent Besides the T-cell response that needs direct cell-cell contact
between IS effector cells and infected cells, the IS uses cells that -- after
having been activated -- secrete
molecules which may bind to the virus particles to mark them for later 
destruction. These cells are called B-cells and the marker molecules they
produce are called \emph{antibodies} (\emph{AB}). Similar to the T-cells, 
the B-cells exist
in our simulation as a pre-activation type \emph{NB} (naive B-cell) and as 
activated B-cells of type \emph{B}.\\
The \emph{NB} cells express on their surface B-cell receptors called \emph{BCR}. The
epitopes of these receptors are randomly determined -- analogously to the
random epitope selection of \emph{NT} cells. Some of the \emph{NB} cells may possess
\emph{BCR} molecules on their surface that are able to bind to the virus \emph{V}
while others may possess receptors that bind to molecules that 
are used by the cells of the IS. The \emph{NB} cells possess mechanisms that
make them present the epitopes of anything which got bound by their \emph{BCR} 
to the extracelular milieu with the help of membrane molecules called \emph{MHC2}. 
If a naive B-cell encounters a T-helper cell \emph{TH} that has \emph{TCR} 
which bind to the \emph{MHC2} of the B-cell (including the additional epitopes
the \emph{MHC2} presents) the B-cell gets activated.
Activated B-cells start secreting antibodies AB that have the same 
random epitopes that were used by the
\emph{BCR}s of the naive cell. These antibodies are hence capable of binding to the
virus that activated the T-helpers.  
The viruses that are marked for destruction by \emph{AB}s are removed
by cells of type \emph{macrophage}. These cells have receptors that 
fit to a non-random
binding site of antibodies. In this way they are able to destroy the viruses
that are bound to antibodies.\\

\begin{figure}[h]
\begin{center}
\includegraphics[width=8cm,angle=270,clip=]{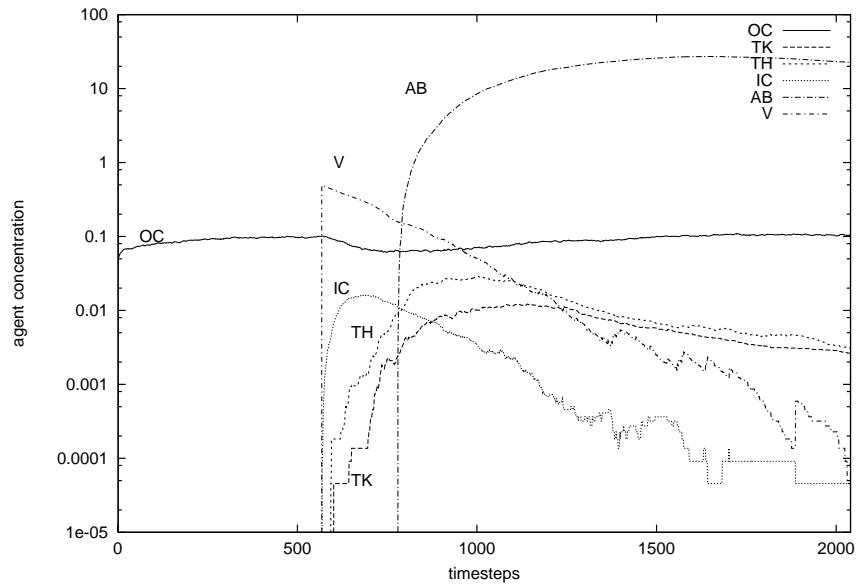}
\end{center}
\caption{time dependency of the concentration of 
organism cells (OC), T-killer cells (TK), T-helper cells (TH), infected cells
(IC), antibodies (AB) and virus (V) in the simple IS model.}
\label{simsim}
\end{figure}

\noindent Fig. \ref{simsim} shows the behavior of this simplified IS. The system starts
with a certain concentration of \emph{OC} that quickly enters a plateau
concentration. Then the virus \emph{V} is injected at a high dose and infection spreads:
the concentration of infected cells \emph{IC} grows. After a while, infected
cells meet \emph{NT} cells that possess the appropriate receptors
to react to the \emph{MHC} presenting the virus-epitope. T-helpers and -killers
appear. T-cell proliferation stops at a certain concentration of T-cells 
as the encounters between T-cells and thus the \emph{FAS:FAS-ligand} induced 
T-cell death get frequent.

B-cells, once being activated by T-helpers, start producing \emph{AB}.
As the T-cell concentration grows, infected cells are effectively removed
from the system. High antibody concentration allows the removal of so many
viruses that the rate of infection of \emph{OC} finally gets low enough to let
the organism experience reconvalescence -
the \emph{OC} concentration returns to its (contact-inhibition controlled)
plateau. As in reality, the antibodies in our simulation not only
mark viruses for destruction, they also may block the binding sites of the viruses
and prevent them from attaching themselves to their target cells \emph{IC}. 

\subsubsection{B-Cell Activation}
B-cells in real ISs may be activated in two different ways. One of them depends on
the assistance of T-helper cells, but B-cells are also capable of responding 
to certain kinds of antigens directly, i.e. without the assistance of
T-cells. These antigens need to possess
several indentical epitopes at the right distance from each other
as to allow a simultaneous binding of more than one B-cell antigen receptor 
(BCR). Polysaccharides as they appear on the membranes of bacteria have this
property. The aggregation of several antigen receptors on the membrane 
of a B-cell triggers a cascade of intracellular mechanisms of the B-cell that 
leads to its activation. The cell will differentiate into an antibody secreting
cell and will proliferate.\\
This T-cell independent B-cell activation features an interesting 
dose-response curve. Instead of inducing a stronger response, very
large concentrations of an antigen lead to an attenuated B-cell activation.
The reason for this is that if the antigen is present in abundance, the
probability is rather high that each of the receptors gets bound by
an antigen molecule of its own. The probability of B-cell receptor
aggregation decreases. (For a mathematical discussion of this effect
as well as for a list of references to work on real ISs see \cite{PERELSON}.)\\
In our computer-experiment we inject a large dose of antigen (\emph{AG}) 
concentrated at one of the walls of the compartment.
While the antigen diffuses we investigate the B-cell activation
by looking at the concentration of the molecules \emph{A} that are produced by
the activated B-cells in slices of the compartment parallel to the gradient of
the antigen concentration.
\begin{figure}[h]
\begin{center}
\includegraphics[bbllx=214,bblly=129,bburx=570,bbury=815,width=6cm,angle=90,clip=]{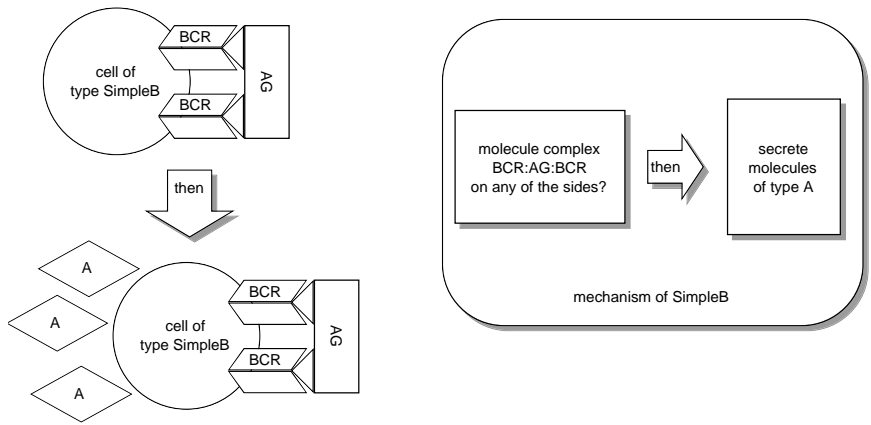}
\end{center}
\caption{mechanism of simple B-cell: cross linking
of two B-cell receptors makes the B-cell secrete \emph{A}.}
\label{simpleb}
\end{figure}

In the simulation simple B-cells are created with the ability to distinguish
signals from six different sides. They possess one important mechanism which
is illustrated in Fig. \ref{simpleb}.
\begin{figure}[h]
\begin{center}
\includegraphics[bbllx=55,bblly=47,bburx=541,bbury=776,width=6cm,angle=270,clip=]{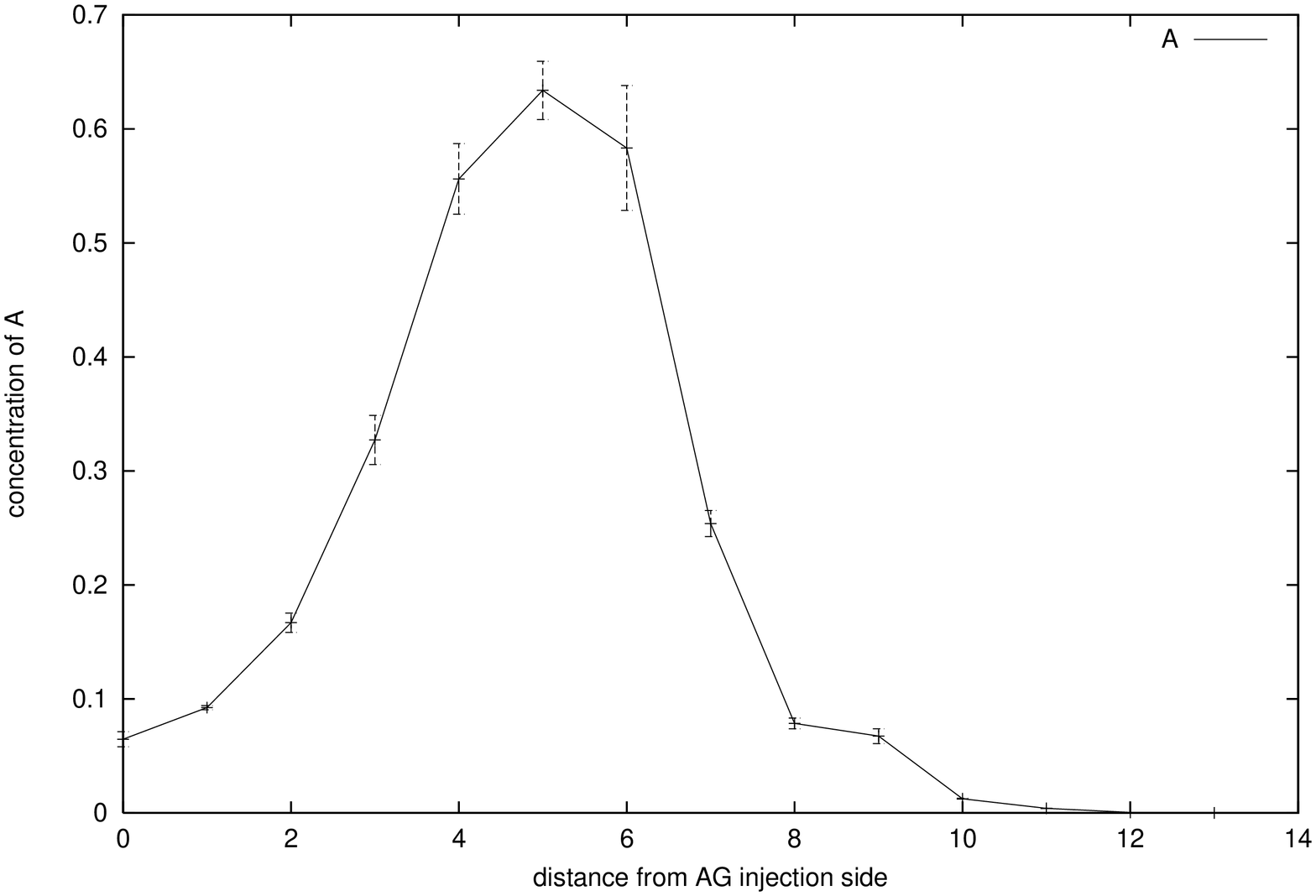}
\includegraphics[bbllx=55,bblly=47,bburx=541,bbury=776,width=6cm,angle=270,clip=]{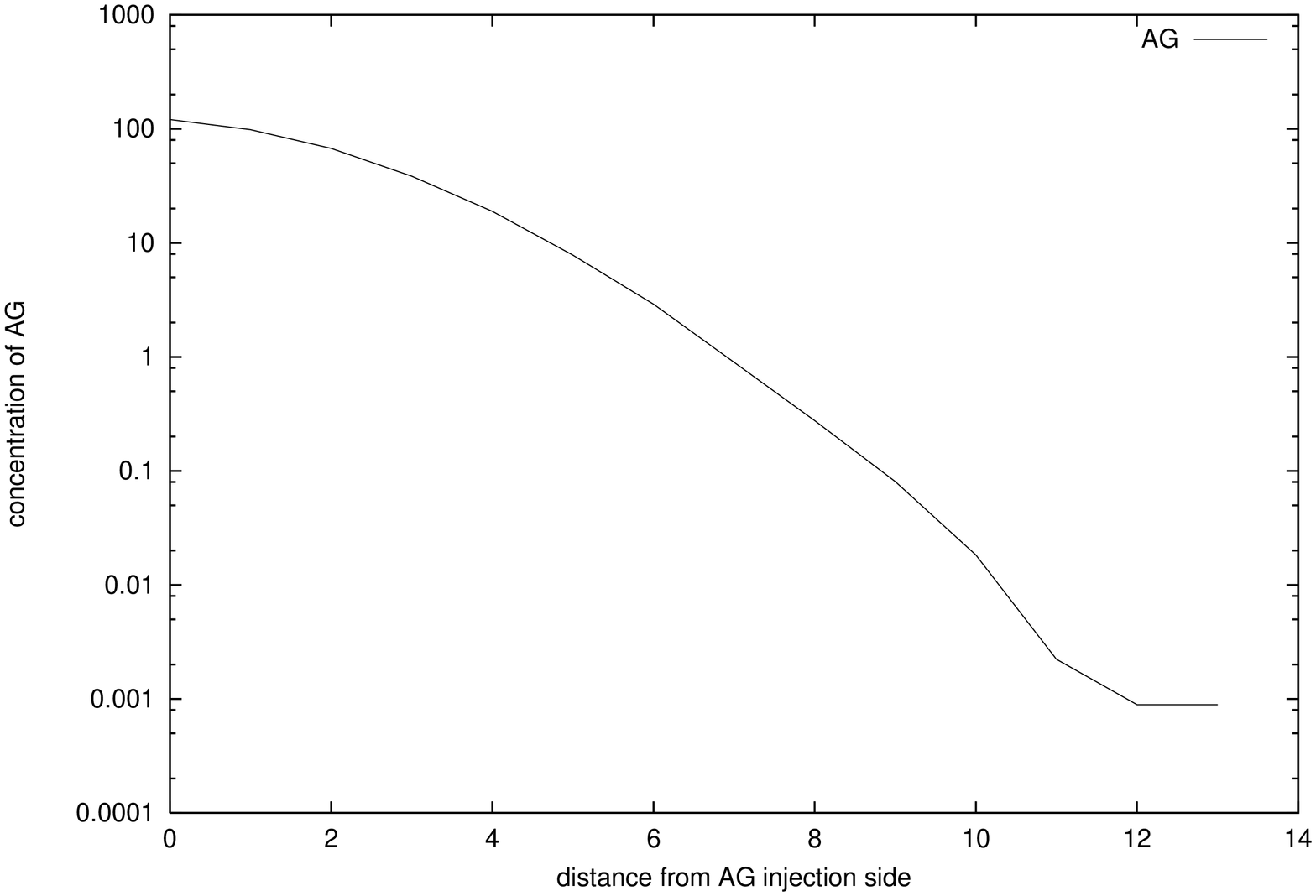}
\end{center}
\caption{Spatial variation of the concentration of molecules
  \emph{A} produced by activated B-cells and of the concentration of
  activating antigen \emph{AG}.}
\label{aggcurve}
\end{figure}

The binding of the \emph{AG}-epitopes by the receptors is 
reversible (with equal reaction constants for binding and release of \emph{AG}) 
and the \emph{A} molecules desintegrate after a certain mean
lifetime. The \emph{AG} molecules are stable. The resulting concentration of
\emph{A} thus indicates how successfully receptor-antigen-receptor aggregates 
are produced.
Fig. \ref{aggcurve} shows the number of \emph{AG} and \emph{A} molecules per cell 
as a function of the distance from the wall of the compartment where the high
antigen concentration was initially injected. The maximum of the concentration
of \emph{A} is located around the area where
10 \emph{AG} molecules per cell can be found. This corresponds -- as expected
-- to just below 2 \emph{AG} molecules per side of the cell.

\subsection{Simulation Analysis}
The main purpose of SIMMUNE is to provide a tool
to investigate how context adaptive behavior of the IS might emerge from
local cell-cell and cell-molecule interactions.\\
The most obvious kind of context dependent behavior of the IS can be found 
in the ability of single cells to react differently to a stimulus when it comes
in combination with other stimuli as opposed to an isolated event.
Thus, mechanisms with several conditons already encode context dependent
behavior.\\
More difficult is the analysis of context recognition that emerges from
cooperations between several cells.
Cooperations between cells manifest themselves as correlations between the 
actions of these cells. A very primitive example from the simulation of section
{\bf 3.3.2} would be the correlation between the \emph{'kill'}-action of T-killer
cells and the \emph{'die'}-action of infected cells.\\
Whether a cell-cell interaction is of just local importance or part of a
network of interdependent cell actions cannot be decided on the microscopical level
of single cells. It requires to analyze on multiple scales the correlations 
between cell actions.\\
Analyzing the behavior of a system on multiple scales means identifying on
each scale collections of objects that may be regarded as \emph{single}
composite objects on a coarser scale. This identification of composite objects
is based on the recognition of common attributes or coherent behavior of the
component objects. In order to allow an analysis that involves a series of
scales that emerge from each other, the rules for the composition of objects
must be applicable not only to the elementary objects of the original, finest
scale but also to the composite objects of coarser scales.\\
In our case, the elementary objects are cell actions. Collections of them are 
identified to constitute composite objects if they are spatially and
temporally correlated. Iterating the process by applying the correlation
analysis to the objects of the first coarse scale may yield a further
scale that is coarser than the first in two aspects: The components of its
objects are themselves composite objects and the correlations between the
constituents of its objects involve larger spatial and temporal distances.\\
Starting from single-cell behavior we enter the scale of cell-cell
cooperations by looking for correlations between cell actions. Looking for
correlations between cell cooperations takes us to the next scale.
Repeated application of this process might lead to a scale that directly
describes the macroscopical behavior of the IS.\\
The network of interactions of our example simulation from section {\bf 3.3.2}
leads to two (yet simple) coarse scales: On the first we register the
correlations between the actions of IS effector cells like T-killer cells and
non-IS cells like infected cells. The second coarse scale contains for example
the correlation between T-cell activation and the appearance of activated
B-cells.\\
Having explained how we proceed in order to look for context 
dependent behavior of the IS we are now able to give a definition of 
what we call a 'recognized immunological context': An immunological context that is
recognized by the IS\nopagebreak \footnote{Context dependent behavior of single cells as
  we mentioned above would be encoded in cellular mechanisms. We want to 
distinguish between such context recognition and the context recognition that
emerges from multi-cellular cooperation.}
 leads to a response of the IS that can be traced on
multiple scales as described above. It can be distinguished from other
contexts that are descibed by different patterns of correlation.\\

Recently, the principle of identifying biological contexts on different
scales by iterated clustering of data has successfully been applied to the
infrared analysis of human blood serum by Werner et al. \cite{GW}.\\

Usually, models about higher level cooperative effects within the IS (like 
automata models that describe the behavior of the IS in terms of predefined
states) must be based on assumptions that try to bridge the scale gap between
the scale of direct cellular interactions and overall behavior of the IS.\\
We do not need to make such assumptions. Cooperative phenomena are detected
using methods of statistical analysis.\\
Of course we pay a price for this: Since SIMMUNE is entirely based on direct
cell-cell or cell-molecule interactions we need to make assumptions about
cellular mechanisms whenever we make use of aspects of single-cell behavior
that are not yet understood in detail.
 
\subsection{Limitations of SIMMUNE}
Due to the faithful modelling of cell behavior, computer power requirements
of SIMMUNE are quite high. With currently available single processor
computer power and memory the maximum number of cells (per computer processor)
 that can participate in a simulation is practically not larger than \emph{500.000}.\\
Recently Smith et al. \cite{SMI} presented a technique
called \emph{lazy evaluation} to simulate quite realistic clone numbers and
sizes. \emph{Lazy evaluation} makes use of the fact that out of the large
number of clones only a small fraction is actually able to interact with a 
given antigen and the presenting cells of the IS. Only this fraction will be 
activated to proliferate and play an important role in the immune response.
By simulating only this fraction while limiting its overall concentration
to the value it would have with the full cell repertoire being existent,
major memory and cpu-time savings can be achieved. To accelerate the
simulation of those models that do not need to consider idiotypic networks
(see \cite{JERNE} or \emph{e.g.} \cite{PERELSON})
future versions of SIMMUNE may adopt this technique.\\
Another limitation of SIMMUNE will not be circumvented by increasing computer
power or techniques as lazy evaluation: As mentioned in the foregoing section,
simulations with SIMMUNE
sometimes have to involve mechanisms and parameters that have not yet been
established by experiment. This often makes quantitative predictions
difficult. On the other hand, the need for detailed descriptions of
cellular mechanisms may point out gaps in immunological knowledge that 
have not yet been sufficiently investigated.

\subsection{Beyond Immunology}
Since all celltypes are user-defined and the basic cellular actions of 
SIMMUNE are typically as general as 'expression of (user-defined) membrane 
receptors' or 'movement along gradients of concentrations of molecules', 
the software may be used to simulate populations of cells of any kind. It has
already been used to simulate systems of neurons with some of the molecular
structure of synaptic gaps between them.

\section{Conclusion}
Besides providing its user with the possibility to define in detail the
behavior and properties of cells and molecules in simulations of the IS,
the mechanism based approach of SIMMUNE yields the fundament for a new
kind of simulation analysis in immunology. Multiscale correlation analysis
of cellular actions may allow the automated classification of immunological
contexts that have not yet been considered until now.\\
Institutes interested in working with SIMMUNE may contact us to receive a
copy of the program together with a manual.\\

\section*{Acknowledgements}

The authors wish to thank Prof. Irun Cohen for helpful discussions
and for providing copies of his forthcoming book.\\
Dr. Ute Kerres (Univ. of Hamburg) has contributed to this work during many
valuable discussions. We thank Dr. Andrew Yates (Institute of Child Health,
London) for reading the manuscript.\\
Support by Deutsche Forschungsgemeinschaft (GRK 135/3-98) and GIF travel
grants are gratefully acknowledged.

\pagebreak
    
\appendix
\section{Simulation Parameters and Diagrams}

\noindent Parameters for \emph{reaction-kinetics} simulations from {\bf  2.1}:\\

{\flushleft
\begin{tabular}{|l|c c c|c|c c c|c c c|} \hline
figure & $p_{infect}$ & $p_{kill}$ & $p_{resp}$ & $s$ & $d_{I}$ & $d_{K}$ &
$d_{C}$ & ${I}_{0}$ & ${K}_{0}$ & ${C}_{0}$ \\ \hline
Fig. \ref{cro} & 0.3        & 0.5      & 0.1      & 0.01 & 0.01 & 0.01 & 0.01 & 0.1 &
0.1 & 1.0\\ \hline
Fig. \ref{osclog} & 0.3        & 1.0      & 0.8      & 0.01 & 0.01 & 0.01 & 0.01 & 0.1 &
0.1 & 1.0\\ \hline
\end{tabular}
}

\vspace{1cm}
\noindent Parameters for the simulation from {\bf  3.3.1}:\\

{\flushleft
\begin{tabular}{|l|c c c c c c c|} \hline
 & $OC$ & $ID0$ & $ID1$ & $ID2$ & $AID$ & ${C}_{1}$ & ${C}_{2}$ \\ \hline
initial concentration & 0.05        & 0.01      & 0.005      & 0.005 & 0.01  &
0 & 0\\ \hline
av. lifespan (timesteps) & $\infty$ & $\infty$ & $\infty$  & $\infty$ & $\infty$ &
100 & 100\\ \hline
\end{tabular}
}\\

\noindent molecular diffusion rate: 0.0001\\
cellular diffusion rate:  0.01\\
(The diffusion rate defines the per timestep probability for agents to get transported
to a neighbor grid point. In this simulation cellular diffusion replaces
active cell movement. The product of molecular diffusion rate and molecular
lifespan defines the \emph{mean} range of a molecular signal emitted by a cell.)\\
 
\noindent compartment dimensions: 80 x 80 x 10\\
 
\begin{figure}[h]
\begin{center}
\includegraphics[bbllx=124,bblly=46,bburx=594,bbury=842,width=10cm,angle=0,clip=]{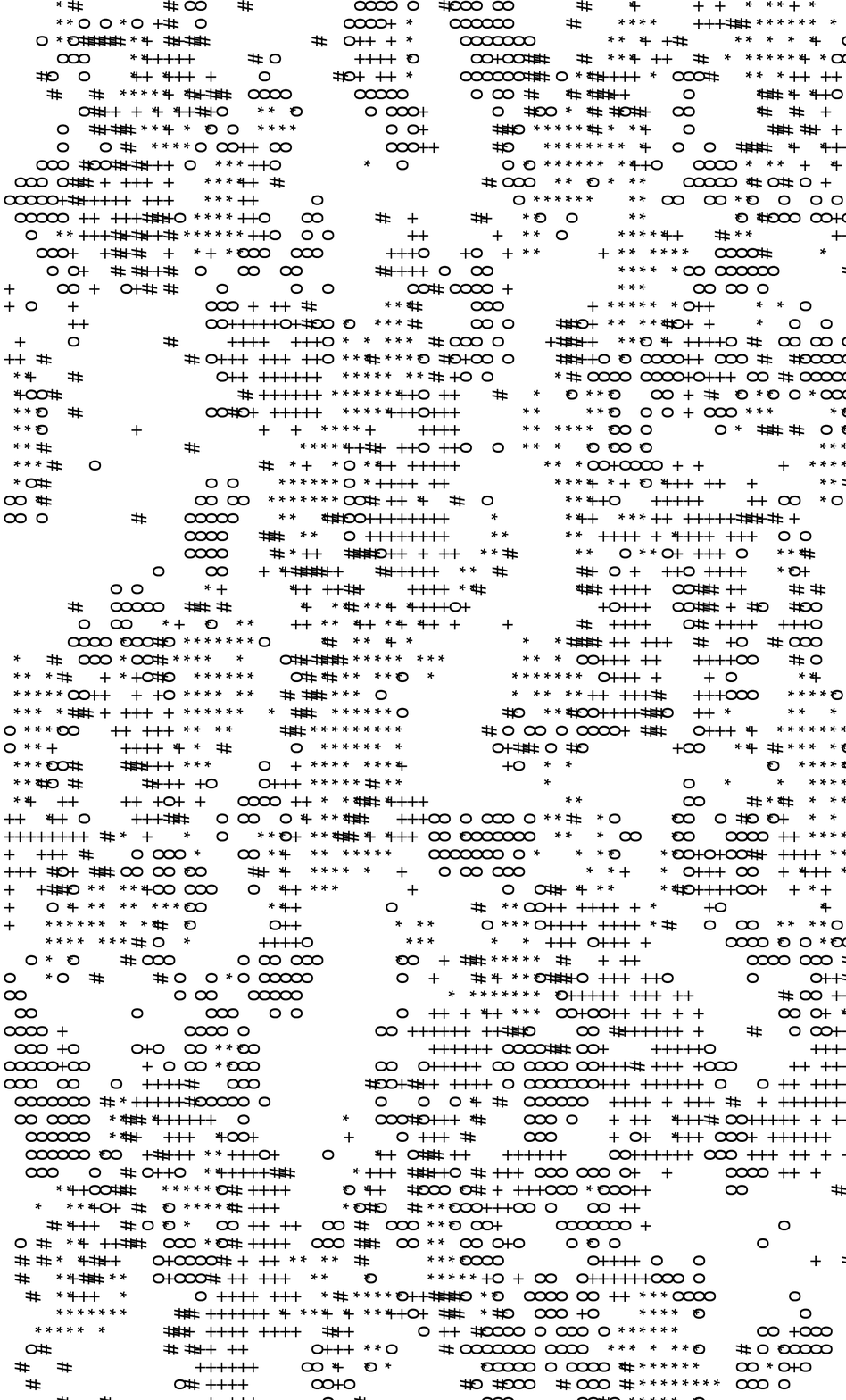}
\end{center}
\caption{clusters in feedback simulation from {\bf  3.3.1}; symbols: 
 $ID0 ~o$, $ID1 ~*$, $ID2 ~+$,$AID ~\#$}
\label{symslice}
\end{figure}

\pagebreak

\end{document}